\documentclass[aps,prl,twocolumn,showpacs,floatfix]{revtex4}

\begin{document}

\title{Energy gaps and interaction blockade in confined quantum systems}

\author{K. Capelle$^{1,2}$,
M. Borgh$^2$, K. K\"arkk\"ainen$^2$ and S.M. Reimann$^2$} 
\affiliation{$\mbox{}^1$Departamento de F\'{\i}sica e Inform\'{a}tica,
Instituto de F\'{\i}sica de S\~ao Carlos, Universidade de S\~ao Paulo,
Caixa Postal 369, 13560-970 S\~ao Carlos, SP, Brazil}
\affiliation{$\mbox{}^2$Mathematical Physics, LTH, Lund University,
22100 Lund, Sweden}
\date{\today}

\begin{abstract}
Many-body effects in confined quantum systems pose a challenging problem 
due to the simultaneous presence of particle-particle interactions and
spatial inhomogeneity. Here we investigate universal properties of strongly
confined particles that turn out to be dramatically different from what is 
observed for electrons in atoms and molecules. We show that for a large 
class of harmonically confined systems, including small quantum dots and 
optically trapped atoms, {\em many-body} particle addition and removal 
energies, and energy gaps, can {\it accurately} be obtained from 
{\em single-particle} eigenvalues. Transport blockade phenomena 
are related to the derivative discontinuity of the exchange-correlation functional. This implies that they occur very generally, with Coulomb blockade being 
a particular realization of a more general phenomenon. In particular, we 
predict {\em van-der-Waals blockade} in cold atom gases in traps. 
\end{abstract}


\maketitle

\newcommand{\be}{\begin{equation}}
\newcommand{\ee}{\end{equation}}
\newcommand{\bea}{\begin{eqnarray}}
\newcommand{\eea}{\end{eqnarray}}
\newcommand{\bi}{\bibitem}
\newcommand{\la}{\langle}
\newcommand{\ra}{\rangle}
\newcommand{\ua}{\uparrow}
\newcommand{\da}{\downarrow}
\renewcommand{\r}{({\bf r})}
\newcommand{\rp}{({\bf r'})}
\newcommand{\rpp}{({\bf r''})}

Many-body effects in confined systems of interacting quantum 
particles are a recurring theme in many areas of science, ranging from 
nuclei over molecules to nanostructured semiconductors~\cite{qdotreview}. 
More recently, much interest turned to the physics of trapped cold atom gases.
With an initial focus on Bose-Einstein condensation~\cite{traprmp}, 
today also  the properties of confined fermionic atoms~\cite{jin} are of 
much concern, and the ability to manipulate trapped atoms recently led to the
suggestion of {\it atomtronics}~\cite{atomtronics}. 
Confinement is often modeled by harmonic-oscillator potentials,
but in spite of decades of research the many-body physics  
in the microscopic as well as the mesoscopic regime is
still not fully understood. Even quantities as fundamental as particle 
addition and removal energies and energy gaps are hard to calculate if 
effects of confinement and of particle-particle interactions are of comparable
magnitude. 

In electronic-structure calculations, addition and removal 
energies and gaps are often calculated from density-functional theory 
(DFT)~\cite{kohnrmp}. 
A large body of knowledge has been accumulated on how such 
calculations should be done, and when their results are reliable. This 
knowledge, however, is largely based on the behavior of electrons in atoms, 
molecules and solids. Extrapolation to other systems is fraught with dangers, 
and may lead astray in many ways. 

In this work, we reassess the calculation of these quantities
in confined systems. A key ingredient of our analysis are near-exact
ground-state energies, obtained from diagonalizing the many-body Hamiltonian, 
which allow an unbiased assessment of approximate schemes. As concrete 
examples, we consider electrons in small quantum dots~\cite{qdotreview} 
and fermionic atoms in optical traps~\cite{jin}. 

Surprisingly, we find that accurate particle addition and removal energies 
can be obtained from local-density single-particle potentials, which
is not at all what one would expect from experience with atoms, molecules and 
bulk semiconductors. From addition and removal energies we calculate energy 
gaps, and estimate the effect of the derivative discontinuity in confined
systems. We relate this discontinuity to Coulomb blockade~\cite{footnote0}, 
which allows us to adopt a more general view on blockade phenomena than the 
usual one, leading, in particular, to the prediction of {\it van-der-Waals 
blockade} in systems of trapped cold atoms.

To begin with, we define a few basic terms and concepts.
The particle-removal energy of an $N$-particle quantum system is defined as
\be
E_r(N) := E(N-1)-E(N) = I(N) = -\epsilon_N(N),
\label{erdef}
\ee
and the particle-addition energy is given by
\be
E_a(N) := E(N)-E(N+1) = A(N) = -\epsilon_{N+1}(N+1).
\label{eadef}
\ee
Here, $E(N)$ denotes the many-body ground-state energy of the $N$-particle
system and $\epsilon_N(M)$ denotes the $N$th eigenvalue of 
the $M$-particle system. When applied to molecules, $E_r(N)$ and $E_a(N)$ 
become the ionization energy, $I$, and electron affinity, $A$, respectively. 
These quantities are defined as differences between ground-state energies 
pertaining to different systems. In DFT, ground-state energies are readily 
calculated 
from the ground-state density, which in turn is obtained from the Kohn-Sham 
(KS) orbitals. Together with these orbitals, one also obtains KS 
eigenvalues. Although these are sometimes taken as a zero-order approximation 
to quasi-particle energies, most of them have no physical significance. The 
exception to this rule is the energy of the highest occupied state, whose 
negative is the ionization energy~\cite{almbladh,lpion,lpsion}. 
The electron affinity can also be obtained from a highest occupied KS 
eigenvalue, albeit that of the $N+1$ particle system.

In approximate calculations of ionization energies of atoms and molecules
by means of the local-density approximation (LDA), or by semi-local 
improvements, it is typically found that values obtained from LDA 
total energies 
agree much better with experiment than do those obtained from eigenvalues. 
The latter can be off by a a huge margin~\cite{footnote1}. For affinities, 
the situation is even worse: the LDA potential often does not even bind one 
additional electron, predicting instead the anion to be 
unstable~\cite{footnote2,trickey,sanvito}. 
Commonly, the bad performance of LDA eigenvalues 
for predicting ionization energies and electron affinities is attributed to 
the erroneous asymptotic decay of the LDA exchange-correlation ($xc$) 
potential. 
For electrons, the exact $xc$ potential (i.e., the one reproducing the exact 
density) decays as $1/r$, whereas the potential corresponding to local 
approximations decays as the density itself, i.e., exponentially. The 
exponential decay of the LDA potential means that the outermost electrons 
are not bound strongly enough, rendering their energies useless for the 
prediction of electron addition and removal energies. Self-interaction 
corrections~\cite{sanvito,pz81} are believed to be required to cure this 
problem. 

The difference between the particle-removal and addition energies
is the fundamental energy gap
\bea
E_g(N) := E_r(N) - E_a(N) 
\label{gapdef1}
\\
= E(N+1)+E(N-1) - 2E(N) 
= E_g^{KS} + \Delta_{xc}
\label{gapdef2}
\\
= \epsilon_{N+1}(N)-\epsilon_N(N) + v_{xc,+}\r - v_{xc,-}\r. 
\label{gapdef3}
\eea
Here $v_{xc,\pm}\r=\delta E_{xc}[n]/\delta n\r|_{N\pm \eta}$, with $\eta\to0^+$
is the exchange-correlation ($xc$) potential calculated at the particle-rich
and the particle-poor side of integer particle number $N$.
$E_g^{KS} = \epsilon_{N+1}(N) - \epsilon_N(N)$ is the KS gap, and
$\Delta_{xc}=E_g-E_g^{KS}$ is known as the derivative discontinuity, because
in DFT it obeys $\Delta_{xc}=v_{xc,+} - v_{xc,-}$ \cite{lpdisc,ssdisc}.
$\Delta_{xc}$ describes a gap that opens upon addition of a single particle
to the system, but disappears in the absence of interactions. Because of these
characteristics it is, in solids, identified with the Mott 
gap~\cite{lpdisc,ssdisc,mottepl}.

In quantum dots, the energy gap is traditionally also decomposed into two
contributions. The single-particle contribution $\Delta \epsilon$ describes 
the effects of quantized energy levels due to geometry and confinement. The 
charging energy $e^2/C$ is due to many-body effects which discontinuously 
raise the energy gap upon addition of one more electron. This effect is hard 
to describe quantum mechanically, and is therefore typically described 
phenomenologically, by a classical capacitance $C$~\cite{qdotreview,kastner}.

Both decompositions of the full gap must add up to the same value, so that
$E_g^{KS}+\Delta_{xc}=\Delta \epsilon+e^2/C$. In the phenomenological approach,
no general microscopic expressions for $\Delta \epsilon$ and $C$ are given.
If $\Delta \epsilon$ is calculated from eigenvalues of noninteracting
particles, subject only to the confining potentials, $e^2/C$ accounts for
all many-body effects and is the only $N$-dependent contribution. DFT suggests 
the alternative identification of $\Delta \epsilon$ with the KS gap, and 
$e^2/C$ with $\Delta_{xc}$. In this case, both $\Delta \epsilon$ and $e^2/C$
depend on interaction and particle number, because $E_g^{KS}$ is obtained
from the eigenvalues of an effective potential, containing selfconsistent
Hartree and $xc$ terms. The alternative identification has the advantage 
that $E_g^{KS}$ is routinely obtained from DFT codes. The capacitance $C$
then describes the beyond-mean-field contribution to blockade, which need
not vanish for $N \gg 1$ ~\cite{footnote3}.

Common local and semi-local density functionals do not have a 
discontinuity, and their prediction for the many-body fundamental gap is 
the KS gap, which can be wrong by a large margin. In principle, however, 
$\Delta_{xc}$ can be estimated from eigenvalues obtained by separate 
calculation of two different systems~\cite{flavia}: combining 
the right-hand sides of Eq. (\ref{erdef}) and (\ref{eadef}) with 
Eq. (\ref{gapdef1}), and comparing the result with Eq. (\ref{gapdef2}), one
finds
\be
\Delta_{xc}=\epsilon_{N+1}(N+1)-\epsilon_{N+1}(N),
\label{flaviaeq}
\ee
which allows one to estimate $\Delta_{xc}$ (defined as a contribution
to the many-body gap) even in situations where the functional used to generate
the eigenvalues has no discontinuity.
Below, we calculate the different contributions to the many-body gap, as
well as separate discontinuities and electron addition and removal energies,
for electrons in quantum dots as well as for 
harmonically confined atoms in optical traps.

First, we turn to quantum dots. Here, we consider Coulomb-interacting 
electrons, which we treat by exact diagonalization and, separately, by the 
local-density approximation to DFT, in the two-dimensional parameterization 
of Attaccalite {\it et al.}~\cite{amgb}. 
Results for electron-removal energies, electron addition energies and
energy gaps are summarized in Tables \ref{table1} and \ref{table2}.

\begin{table}
\caption{\label{table1}Negative of the electron-removal energies of the 
$N$-particle dot, obtained exactly ($E_r^{MB,\Delta E}$), from LDA 
total-energy differences ($E_r^{LDA,\Delta E}$), and from LDA eigenvalues 
($E_r^{LDA,ev}$). These are also the negative electron-addition energies 
of the $N-1$-particle systems. (Here and below $\Delta E$ and $ev$ refer to 
calculations as total-energy differences and from eigenvalues, respectively.)
All values are in atomic units.}
\begin{center} 
\begin{tabular}{l|l|c|c|c}
$N$ & $\omega$ & $-E_r^{MB,\Delta E}$ & $-E_r^{LDA,\Delta E}$ & 
$-E_r^{LDA,ev}$ \\
\hline 
5&0.15&1.16&1.16&1.24\\
 &0.25&1.68&1.68&1.79\\
 &0.35&2.14&2.15&2.28\\
6&0.15&1.35&1.33&1.41\\
 &0.25&1.92&1.91&2.02\\
 &0.35&2.44&2.43&2.56\\
7&0.15&1.60&1.55&1.62\\
 &0.25&2.30&2.24&2.34\\
 &0.35&2.92&2.87&2.99\\
\end{tabular}
\end{center}
\end{table}

Several aspects of these results are surprising, and unexpected from 
experience with atoms, molecules and bulk semiconductors. First, we note,
in Table \ref{table1}, that electron addition and removal energies obtained 
from eigenvalues are quite close to the exact data. For larger particle 
numbers, addition and removal energies obtained from LDA eigenvalues can 
even be better than those from total-energy differences.

This behavior of addition and removal energies obtained from eigenvalues is 
dramatically different from a huge body of experience accumulated for atoms, 
molecules and solids, where eigenvalue-based ionization energies
differ widely from experiment~\cite{footnote1}, and affinities sometimes 
cannot be obtained at all~\cite{footnote2}. This difference shows that 
much of the problems commonly associated with the LDA, or with LDA KS 
eigenvalues, are not really due to the LDA itself. Rather, the explanation 
must lie in a physical difference between harmonically confined fermions and 
molecular systems.  

\begin{table}
\caption{\label{table2}Energy gaps of quantum dots, multiplied by 10 for
legibility. Seem main text for explanation of symbols.}
\begin{center}
\begin{tabular}{l|l|c|c|c|c|c}
$N$ & $\omega$ & $E_g^{MB,\Delta E}$ & $E_g^{LDA,\Delta E}$ & $E_g^{LDA,KS}$
& $\Delta_{xc}$ & $E_g^{LDA}$\\
\hline
4&0.15&2.16&2.29&0.547&1.60&2.15\\
 &0.25&3.21&3.23&0.862&2.20&3.06\\
 &0.35&3.88&3.99&1.07&2.69&3.76\\
5&0.15&1.98&1.70&0.110&1.67&1.78\\
 &0.25&2.38&2.30&0.122&2.15&2.28\\
 &0.35&3.00&2.82&0.127&2.66&2.78\\
6&0.15&2.46&2.16&0.654&1.46&2.12\\
 &0.25&3.76&3.30&1.22&2.00&3.22\\
 &0.35&4.80&4.39&1.84&2.45&4.29\\
\end{tabular}
\end{center}
\end{table}

In particular, and in contrast to common physics lore, this explanation cannot
be related to the asymptotic decay of the LDA potential: For harmonic 
confinement of Coulomb-interacting particles, the $1/r$ behavior of the 
exact $xc$ potential is still observed, both in three~\cite{burke} and 
two~\cite{damico} dimensions, but the LDA single-particle orbitals and density 
now decay as Gaussians. As Fig. 24 of Ref.~\cite{qdotreview} shows, the 
self-consistent LDA density agrees well with the many-body density, even in 
the asymptotic region. As a consequence, $v_{xc}^{LDA}\r$ 
now also decays as a Gaussian, i.e., even faster than exponentially, which 
could be expected to worsen the performance of the LDA eigenvalues, instead 
of improving it.

The effective potential, however, is much stronger than for atoms,
because it contains the external confining potential. In a harmonic potential, 
the system is completely confined, i.e., any number of electrons is bound and 
there are no continuum states. The absence of continuum states in 
harmonic confinement, which is a realistic feature of real quantum dots at 
low energies and atoms in optical traps, is behind the improved binding of 
the anion-like states, and thus, by means of Eqs.~(\ref{erdef}) to 
(\ref{flaviaeq}), also behind the other improvements noted. The erroneous
asymptotics of the LDA potential are not sampled by the confined particles.

It is also noteworthy that standard proofs \cite{almbladh,lpion,lpsion}
of the identification $-\epsilon_N(N)=I$ all explicitly or implicitly assume 
that the external potential decays to zero as $|{\bf r}| \to \infty$, or the 
closely related fact that the single-particle orbitals far away from a finite
system decay exponentially with an energy-dependent exponent. Neither is
true for harmonic confinement, which grows indefinitely as $|{\bf r}| \to
\infty$, and produces single-particle orbitals that decay as Gaussians
with a {\em universal} (energy independent) exponent. A generalization of 
the proof to harmonic confinement has been sketched in the appendix of
Ref.~\cite{lpcomment}, and is vindicated by our numerical results.

A second aspect of the data that deserves further investigation is the
behavior of the gaps in Table \ref{table2}. The LDA KS gap $E_g^{LDA,KS}$, 
obtained from eigenvalues of an occupied and an unoccupied orbital, greatly 
underestimates the many-body gap $E_g^{MB,\Delta E}$. This shows that the 
underestimate arises from the lowest unoccupied orbital, not from any occupied 
orbital. On the other hand, energy gaps $E_g^{LDA}$ obtained by adding 
$\Delta_{xc}$ to $E_g^{LDA,KS}$ (or, equivalently, from differences of 
eigenvalues pertaining to the $N$-particle and $N+1$-particle KS systems) 
are slightly different from LDA total-energy differences, $E_g^{LDA,\Delta E}$, 
being sometimes a bit better and sometimes a bit worse, when compared to
$E_g^{MB,\Delta E}$. To within this small 
fluctuation, many-body gaps of quantum dots can thus be obtained from just 
two self-consistent calculations of eigenvalues, instead of from three 
self-consistent calculations of total energies. Note that in these 
calculations the derivative discontinuity, estimated from Eq.~(\ref{flaviaeq}), 
makes a significant contribution to the many-body gap. 

Next, we turn to optically confined atom gases.
The basic formalism of DFT applies both to electrons in quantum dots,
and cold atoms in traps. The functionals, however, are different because 
electrons interact via the Coulomb interaction, whereas atoms interact 
via the van-der-Waals force. In dilute systems of cold atoms, this
is commonly modeled by a suitable contact interaction. Again assuming 
harmonic external confinement, we perform a Hartree calculation, 
$v_{xc}\equiv 0$. The data in Table \ref{table3} show that the atom-removal 
energies obtained from eigenvalues and from total-energy differences are in 
similar good agreement for trapped atoms as for electrons in quantum dots, 
indicating that the key is indeed not the asymptotic behavior of the $xc$ 
potential, but confinement.

\begin{table}
\caption{\label{table3}Negative of the atom-removal energies of
$N$ atoms in a harmonic trap, interacting with a contact interaction
of strength $g$, calculated as total-energy differences and from
single-particle eigenvalues, all obtained in the Hartree approximation.
These are also the negative atom-addition energies of the $N-1$-particle
systems. $E_g^{H}$ is the Hartree single-particle gap, and $\Delta_{xc}$
is the derivative discontinuity, estimated from single-particle eigenvalues
according to Eq.~(\ref{flaviaeq}). At $N=6$ and $g=1.0$ the exact (CI) gap
is $0.9376$, which agrees well with the sum $E_g^{H}+\Delta_{xc}=0.9181$.}
\begin{center}
\begin{tabular}{l|l|c|c|c|c}
$N$ & $g$ & $-E_r^{H,\Delta E}$ & $-E_r^{H,ev}$ &
$E_g^{H}$ & $\Delta_{xc}$ \\
\hline
 6&1.0&2.4172&2.4328&0.8694&0.0487\\
  &2.0&2.7748&2.8009&0.7776&0.0902\\
  &5.0&3.6457&3.6913&0.6113&0.1871\\
10&1.0&3.4528&3.4911&$4.3\times 10^{-4}$&0.0453\\
  &2.0&3.8537&3.9218&$7.8\times 10^{-4}$&0.0804\\
  &5.0&4.8632&4.9963&$1.6\times 10^{-4}$&0.1569\\
\end{tabular}
\end{center}
\end{table}

For the trapped atoms, estimates of the derivative discontinuity are also
represented in Table \ref{table3}. Not unexpectedly, the discontinuity
arising from the van-der-Waals interaction is smaller than that arising
from the much stronger Coulomb interaction, but still makes a significant
contribution to the gap, in particular for larger values of the interaction
parameter $g$. 

Based on the similarity to the quantum-dot case, we predict that a blockade 
phenomenon will also occur if repulsively interacting atoms are channeled 
one by one through an optical trap loaded with a small number of particles 
and coupled to a reservoir \cite{recati}, even if the interaction between 
the particles is of the van-der-Waals type. Interaction-driven blockade is 
not limited to the Coulomb interaction. {\em Van-der-Waals blockade} is 
expected to play a key role in transport experiments on confined cold atoms 
\cite{recati}, and in {\em atomtronic} devices of the type proposed in 
Ref.~\cite{atomtronics}.

In conclusion, an unexpected, but very favorable scenario emerges from this 
analysis: Many-body particle addition and removal energies of confined systems
can be reliably estimated from single-body energies, and the many-body gap and 
its Coulomb-blockade contribution can be obtained with relative ease and good 
precision, even from LDA. These features are expected to apply universally to 
fully confined systems. 

Derivative discontinuities give rise to blockade phenomena, which 
we expect to be ubiquitous, the Mott insulator, Coulomb blockade and 
van-der-Waals blockade being just three particular realizations of a very 
general phenomenon.

This work was financially supported by the Swedish Research Council, 
the Swedish Foundation for Strategic Research, the Finnish Academy 
of Science, and the European Community project ULTRA-1D 
(NMP4-CT-2003-505457). KC is supported by FAPESP and CNPq, and thanks 
L.~N.~Oliveira for useful discussions.


\end{document}